\documentstyle[eqsecnum,preprint,aps,prd,epsf,rotate]{revtex}
\preprint{DAMTP R-98/18}
\date{\today}
\draft
\tighten
\begin{document}
\def\sqr#1#2{{\vcenter{\hrule height.3pt
      \hbox{\vrule width.3pt height#2pt  \kern#1pt
         \vrule width.3pt}  \hrule height.3pt}}}
\def\square{\mathchoice{\sqr67\,}{\sqr67\,}\sqr{3}{3.5}\sqr{3}{3.5}}
\def\today{\ifcase\month\or
  January\or February\or March\or April\or May\or June\or July\or
  August\or September\or October\or November\or December\fi
  \space\number\day, \number\year}

\def\Bbb{\bf}


\title{Existence of Majorana fermions for M-branes \\
        wrapped in space and time}
\author{\sc Andrew Chamblin}

\address {\qquad\\ DAMTP, Silver Street\\
Cambridge, CB3 9EW, England
}

\maketitle

\begin{abstract}
{We show that it is possible to define Majorana (s)pinor fields on M-branes
which have been identified under the action of the antipodal map on the
adS factor of the throat geometry, or which have been wrapped on two-cycles
of arbitrary genus.  This is an important consistency check,
since it means that one may still take the generators of supertranslations
in superspace to transform as Majorana fermions under the adjoint action
of $Spin(10,1)$, even though the antipodally identified M2-brane is {\it not} 
space-orientable.  We point out that similar conclusions hold for any p-branes
which have the generic (adS)$~{\times}~$(Sphere) throat geometry.}  
\end{abstract}
\pagebreak

\section{Introduction}

Recently, there has been an enormous amount of interest in the information
which may be contained in the near-horizon geometries of the p-branes which are
a staple feature of the supergravity and string theory diet.  In particular,
it has recently been conjectured \cite{juan} that information about the
dynamics of superconformal field theories (in the large $N$ limit)
may be obtained by studying the region near the horizon of certain
D(p)-branes.  Thus, the conjecture implies a correspondence between
gauge theories in the large $N$ limit and compactifications of supergravity
theories.  The correspondence is often called `holographic' \cite{ed} because the
superconformal field theory (SCFT) lives on the causal boundary of adS.  This
boundary is the `horosphere' at infinity \cite{gazza} - it is a timelike
hypersurface with the topology $S^{1}~{\times}S^{p}$, where the circle
$S^1$ is the timelike factor. 

Given this correspondence, one may search for new and interesting properties
of SCFTs simply by investigating p-branes with unusual asymptotics.
That is, one may consider p-branes where the throat geometry has been
modified in some way.  One obvious way to modify a given solution, 
is to identify the solution under the action of some freely acting discrete
transformation group.  Recently, Gibbons \cite{gazza} has argued
that such identifications may in fact be necessary in order to avoid
fixed point singularities, in situations where one is `wrapping' a p-brane
on a toroidal cycle.  In particular, he argues that one must compose
any wrapping identifications with the antipodal map on the adS factor of
the near-horizon geometry.  Since the antipodal map is freely acting, the
composition will be freely acting and the resulting identified brane
will be free of fixed-point singularities.

Of course, whenever one identifies a manifold under the action of some
freely acting involution, the resulting manifold may or may not be orientable.
When the identified manifold is {\it non}-orientable, one has to be 
careful to check for the existence of fermions.  That is, one needs to make
sure that there exists a pin bundle with the right properties, so that
any required fermionic fields can exist as sections of the bundle.

In this paper, we check that Majorana pin structure always exists for
M-branes which are identified under the action of the antipodal map on the
adS factor of the near-horizon geometry.  Thus, wrapping M-branes in this
way is not obstructed by the requirement of Majorana pin structure.
We point out that similar considerations
will hold for p-branes in any dimension, as long as the choice of representation
for the parity inversion operator satisfies certain constraints.
Finally, we conclude with some general remarks about the uniqueness of
eleven dimensional SUGRA, and how M-theory may solve the old problem of 
the classification of fermions.

\section{Majorana pinors and wrapping branes}

We are working in eleven dimensions with the convention that the spacetime
has signature $(- + + + + + + + + + +)$.  $D = 11$, $N = 1$ supergravity
is a theory which describes the interaction of gravity with a Majorana
gravitino ${\Psi}_A$ and a three-index gauge field $A_{LMP}$.
The theory has several continuous symmetries: Local $N = 1$ SUSY,
$D = 11$ general covariance, Abelian gauge invariance for the three-form
$A_{LMP}$ and of course ${\mbox{SO}(10,1)}$ Lorentz invariance.  It also has
a discrete symmetry associated with the effect of spacetime reflections
on the gauge field.  This symmetry tells us \cite{duff} that the action 
and equations of motion in eleven dimensions are invariant under an
odd number of spatial (or temporal) reflections, together with the 
reversal of the sign of the gauge field:
\[
A_{LMP} ~{\longrightarrow}~ - A_{LMP}
\]
 
In fact, this discrete symmetry is {\it essential}
whenever we consider non-orientable spacetime manifolds in M-theory.  This is
because we typically think of the four-form $F_{LMNP}$ as being proportional
to some volume form, or anti-symmetric tensor ${\epsilon}_{LMNP}$.
It follows that on an non-orientable manifold, $F_{LMNP}$ will not
have a definite sign - the sign will change when we propagate around a 
non-orientable loop.  However, propagation around an orientation reversing
loop also reflects everything through an odd number of spacetime dimensions,
i.e., the equations of motion are still invariant even though the 
four-form is reduced to the status of a `pseudo-tensor'.  This means
that it still makes sense to talk about the eleven dimensional supergravity
equations of motion on non-orientable spacetimes.  For a further discussion
of non-orientable configurations in M-theory, the reader should 
consult \cite{flux}.

Now, a key thing to notice is that it really is {\it not possible}
to consistently modify this structure in any way.  In particular, the 
Majorana condition for the gravitino is precisely what one needs in order
to match the number of bosonic and fermionic degrees of freedom.
One cannot just flippantly introduce other representations for the fermions.  

A pleasant feature of life in eleven dimensions is the fact that the real
Clifford algebra may be written as
\[
Cliff(10,1;{\Bbb R}) = {\Bbb R}(32)
\]

\noindent ${\Bbb R}(32)$ denotes the space of real 32$~{\times}~$32
matrices and $Cliff(10,1;{\Bbb R})$ denotes the set of objects ${\gamma}_{\nu}$
which satisfy the relation
\begin{equation}
{\gamma}_{\mu}{\gamma}_{\nu} + {\gamma}_{\nu}{\gamma}_{\mu} = +2g_{\mu\nu}
\end{equation}

\noindent where $g_{\mu\nu}$ is the metric on eleven dimensional Minkowski
space with the signature prescribed above.  In the usual way, these gamma
matrices act on a 32 dimensional space of Majorana spinors, 
which are real with respect to the relevant charge conjugation operator 
$C_{ij} = -C_{ji}$.  Explicitly, such a spinor is just a 32 component
column ${\psi}_k$, $k = 1, 2, 3, ... 32$.

It is {\it essential} for the contruction of eleven dimensional
supergravity that we are able to define, globally and consistently,
these Majorana fermions in any eleven dimensional spacetime we wish to consider.
Without such spinors, we can have no gravitino field with the right
number of degrees of freedom and similarly we cannot define generators
of supertranslations in superspace which will transform in the right way.
If there is some topological anomaly or obstruction which prevents us
from defining a globally well-defined spin bundle which has Majorana sections,
then the entire structure will collapse.

Of course, up to now the spacetimes considered in most approaches to 
$D = 11$ SUGRA have had fairly trivial topological characteristics.  As an
example, consider the two objects which couple naturally to the three-form
gauge field: The (electric) M2-brane and the (magnetic) M5-brane.  
The global causal structure
of these M-branes is very familiar.  The 2-brane interpolates between
$(adS)_{4} ~{\times}~ S^{7}$ (which is a supersymmetric compactification
of the eleven dimensional theory)
and flat Minkowski spacetime ${\Bbb M}^{11}$.
In a similar way, the 5-brane interpolates between $(adS)_{7} ~{\times} S^{4}$
and ${\Bbb M}^{11}$.  In each case, as one falls down the throat of the brane
one moves into the region where the vacuum is a standard compactification of
the form $(adS)_{p+2} ~{\times}~ S^{11-p-2}$.

Given this picture, that p-branes are just solutions which describe vacuum
interpolation, one is led to several obvious and natural questions.
For instance,
is it possible to find p-branes which interpolate between vacua which are
exotic, or non-standard, compactifications of the supergravity theory?
What happens if we identify the known solutions, such as the M2 and M5 branes,
under the action of some discrete transformations?  Do the resulting 
`orbifold' branes still make sense?

One potential problem with identifying a given solution 
under the action of some freely acting
involution is the fact that the resulting orbifold brane may be non-orientable.
In particular, the existence of fermions on non-orientable spacetimes is a 
subtle problem.  In order to understand this problem, we need to first
recall some elementary facts about the `pin' groups.

Any proposal to quantize gravity via a path integral prescription, which 
includes a sum over manifold topologies, will obviously force us to
consider the effects of non-orientable manifolds.  A non-orientable
manifold has the property that there exist closed loops in the manifold,
such that parallel propagation around a given loop results in a reversal
of some orientation.  Thus, given a non-orientable manifold of 
signature $(p,q)$, it follows that the tangent bundle of the manifold
can at {\it most} be reduced to an ${\mbox{O}(p,q)}$ bundle.  When we introduce
fermions on the manifold, we `lift' the tangent bundle to a bundle
with fibers given as the group which is the double-cover of the tangent bundle
group.  Thus, we need to know what groups are the double-covers of
${\mbox{O}(p,q)}$ in order to understand how to introduce fermions on a non-oriented
space.

The groups which are double-covers of the group ${\mbox{O}(p,q)}$ are called
the {\it pin} groups.  The notation is meant to be humorous:
Just as ${\mbox{Spin}(p,q)}$ double-covers ${\mbox{SO}(p,q)}$, so does 
${\mbox{Pin}(p,q)}$
double-cover ${\mbox{O}(p,q)}$.  For an excellent review of the history of these
things, the interested reader should see \cite{cec}.

In general, there are in fact eight distinct groups which double-cover
${\mbox{O}(p,q)}$.  These different groups correspond to how one may choose to 
represent the discrete transformations of parity inversion (P), 
time reversal (T), and the combination of the two (PT).  More precisely,
since any of these discrete transformations squares to the identity
in the tangent bundle (i.e., $P^2 = +1$ in the tangent bundle), 
it follows that there is a sign ambiguity in the double-cover (i.e.,
$P^2 = {\pm}1$ in the pin bundle).  It follows that there are $2^3$ groups.
Here, we shall use the notation of Dabrowski \cite{dab} and write these
double-covers as shown:
\[
{h^{a, b, c}:~ {\mbox{Pin}}^{a, b, c}(p, q) ~{\longrightarrow}~ O(p, q)}
\]

\noindent with $a, b, c ~{\in}~ {\{}+, -{\}}$. The signs of $a, b$, and $c$
are defined to be the signs of $P^2$, $T^2$ and $(PT)^2$ respectively.
This is all we will need to know about pin groups.

Now, if we are given a manifold which admits a globally well-defined
pin-bundle, with fibers ${\mbox{Pin}}^{a, b, c}(p, q)$, then we shall say
that the manifold admits a ${\mbox{Pin}}^{a, b, c}(p, q)$-structure.
On a given non-orientable manifold, Majorana fermions will be sections of
a bundle corresponding to some ${\mbox{Pin}}^{a, b, c}(p, q)$-structure.
Thus, the existence of Majorana fermions is equivalent to the existence
of the relevant ${\mbox{Pin}}^{a, b, c}(p, q)$-structure.  We therefore need
to understand how topology can obstruct the existence of a given pin
structure.

\section{Obstructions to Majorana pin structures on wrapped M-branes}

The obstructions to Cliffordian pin structures were worked out in
\cite{max}; this work was extended to include the obstructions to 
all pin structures in any dimension and any signature in \cite{me}.
In this short note, we will not go into the details of obstruction 
theory, or how the obstructions are derived.  However, we do need to 
recall a small set of topological invariants in order to even write
the obstructions down.

In order to do this, we first need some minimal notation.  Let $M$
denote the eleven-dimensional spacetime manifold, and $g_L$ the Lorentzian
metric (with signature as above) on $M$.  The obstructions which we will
describe depend on these two basic objects.  Calculating one of these
obstructions amounts to calculating a number which is an element of
the {\it additive} cyclic group ${\Bbb Z}_{2} = (0,1)$.  A given pin
structure will exist if and only if the relevant obstruction vanishes.

An important invariant here is the {\it second Stiefel-Whitney class},
denoted $w_{2}(M)$.  This invariant, which is the obstruction to the 
existence of spin structure on $M$, is an element of the second cohomology
group $H^{2}(M;{\Bbb Z}_{2})$.  That is to say, $w_2$ may be regarded as
a form, which can be evaluated on two-dimensional cycles in $M$
(the elements of $H_{2}(M)$).
If $w_2$ is non-vanishing on a given two-cycle, it follows that there
does not exist spin structure on $M$.  Explicitly, if $w_2$ did not
vanish on some two-cycle, one would find that
there was an anomaly in a given spinor field, as the spinor field was parallel
propagated around on the two-cycle.

Next, we need the {\it first Steifel-Whitney class}, denoted $w_{1}(M)$.
This invariant is the obstruction to the orientability of $M$, i.e., 
$w_1$ vanishes if and only if $M$ is orientable.  As the name suggests,
this invariant is an element of the first cohomology group,
$H^{1}(M;{\Bbb Z}_{2})$.  $w_1 = 1$ on loops , or one-cycles,
in $M$ which are orientation reversing.

On a Lorentzian manifold, the first Steifel-Whitney class decomposes
into two `sub'-classes, which may be regarded as the obstructions to 
space and time orientability seperately.  In particular, there is the
`spacelike' Stiefel-Whitney class, denoted ${w_{1}}^{S}(M;g_{L})$,
which is the obstruction to the orientability of the spacelike sub-bundle
of the tangent bundle, and likewise there is a timelike class,
denoted ${w_{1}}^{T}(M;g_{L})$, which is the obstruction to 
time orientability.  Obviously,
\[
w_{1}(M;g_{L}) = {w_{1}}^{S}(M;g_{L}) + {w_{1}}^{T}(M;g_{L})
\]

\noindent i.e., if you go around a loop and simultaneously reverse both 
the space and time orientations, then the overall orientation of the spacetime
manifold is not reversed.  Throughout this paper we will assume that 
spacetime is at least time orientable; it follows that $M$ is non-orientable
if and only if it is not space orientable.
This is all of the topological information which we shall need.

Now, we need to decide which pin structure corresponds to the Majorana
fermions described above.  Since all of the orbifold branes which we will
consider here will be time orientable but not space orientable, this means
that we need to make a choice about how we are going to represent the 
parity inversion operator.  Our choice, which is the simplest ansatz
that will give a unitary representation of $O(10,1)$, is the Cliffordian
representation:
\begin{equation}
{\cal P} = {\gamma}_{1}{\gamma}_{2}{\gamma}_{3}{\gamma}_{4}
{\gamma}_{5}{\gamma}_{6}{\gamma}_{7}{\gamma}_{8}{\gamma}_{9}{\gamma}_{10}
\end{equation}

\noindent This is the Cliffordian choice in the sense that this is how you
would represent, in the Clifford algebra itself, inversion through all of
the spacelike coordinates simultaneously.  (Note: We could just as easily
take ${\cal P} = {\gamma}_{0}$, as discussed in \cite{us}, \cite{cec}.
It should be obvious - from what we say below - that this would not affect
the obstruction theory).  On the surface this may seem innocuous,
but there are some real subtleties here.  First, the choice (3.1) for ${\cal P}$
forces us to make the corresponding Cliffordian choice for time reversal:
\begin{equation}
{\cal T} = {\gamma}_{0}
\end{equation}

\noindent This is fine; however, we have to {\it decide} whether we
want to represent time reversal using a unitary operator or an
anti-unitary operator.  Explicitly, we have to ask ourselves:
Do we want to just multiply by ${\gamma}_{0}$ and reverse the sign
of $t$ when we apply ${\cal T}$ to a pinor field (this would give
us a unitary operator), or do we also take the charge conjugate of the
field (this would give us an anti-unitary operator)?
Wigner \cite{wigner} argued that we should use an 
{\it anti}-unitary operator to represent time reversal, since then
time reversal would map positive energy states to positive energy states,
in the quantum mechanical Hilbert space.  Thus, in the Wignerian approach
one no longer works with strictly unitary {\it representations} of
${\mbox{O}(10,1)}$; instead, one works with what Wigner called {\it corepresentations},
which are basically just like unitary representations only some of the
operators are allowed to be anti-unitary.  Here, we are not worrying about
these subleties because we are not trying to do quantum mechanics - we are just
looking for some choices of P and T which will move us around in the fibre of
the pin bundle in the appropriate way.  In any event, taking ${\cal T}$ to 
be anti-unitary will not affect the obstruction theory.  All that matters for our
purposes is that ${\cal P}^2 = -1$.

Given all of this, we can now work out the pin structure we are working with
for these Majorana fermions.  Given the signature of spacetime,
one calculates
\[
{\cal P}^2 = -1, {\cal T}^2 = -1, ({\cal P}{\cal T})^2 = -1
\]

\noindent which means that these Majorana pinors require the existence of
a ${\mbox{Pin}}^{-, -, -}(10,1)$-structure.  Actually, there is one final
subtlety here, namely, we could reverse the signs of all of the gamma
matrices simultaneously.  This would change the sign of $({\Gamma}_{11})^2$,
and so we would be working with a ${\mbox{Pin}}^{-, -, +}(10,1)$-structure.
This choice would not affect the obstruction theory, or anything else,
and need not concern us here.

The topological obstruction to the existence of this pin structure was worked
out in \cite{me}.  One refers to the relevant theorem, and finds that there
exists ${\mbox{Pin}}^{-, -, -}(10,1)$-structure on a time orientable
spacetime $(M, g_L)$ if and only if
the below obstruction vanishes:
\begin{equation}
w_{2}(M) + {w_{1}}^S(M;g_{L})~{\smile}~{w_{1}}^S(M;g_{L}) = {\cal O}(M)
\end{equation}

\noindent where ${\smile}$ denotes the cup-product (see, e.g., \cite{munk}).  Thus,
we see that as long as $w_{2}(M) = 0$ and ${w_{1}}^S(M;g_{L}) = 0$, 
${\cal O}(M) = 0$ and so Majorana structure will exist.  Of course,
if $w_{2}(M) = 0$ then $M$ is a spin manifold anyway and, as we have already
pointed out, many of the orbifold branes will {\it not} be space
orientable and hence ${w_{1}}^S(M;g_{L}) {\neq} 0$ in general.  This obstruction
is therefore non-trivial, and has to be checked for each orbifold brane.

Here, we are primarily concerned with M-branes which have been identified
under the action of the antipodal map on the adS factor of the near-horizon
geometry.  Such antipodally identified branes are said to be `wrapped in
space and time' \cite{gazza}, because the antipodal map on adS involves
an identification of the timelike coordinate.  Explicitly, if we write the
metric on $(adS)_{p+2}$ in static coordinates:
\begin{equation}
ds^2 = -{\mbox{cosh}}^{2}{\chi}dt^2 + d{\chi}^2 + 
{\mbox{sinh}}{\chi}d{{\Omega}_{p}}^2
\end{equation}

\noindent where $d{{\Omega}_{p}}^2$ is the round metric on the sphere $S^p$
and $0 {\leq} t {\leq} 2{\pi}$, then
the antipodal map (denoted J) may be written
\[
J: (t, {\chi}, {\Bbb n}) ~{\longrightarrow}~ (t + {\pi}, {\chi}, -{\Bbb n})
\]

Recently, Gibbons \cite{gazza} has introduced this involution and argued that
it may be an essential ingredient in any scenario where one is wrapping a 
p-brane around a toroidal cycle.  Explicitly, he shows that it is not
possible to find {\it any} finite freely generated abelian group (acting
as spatial translations on the coordinates `tangent' to the brane) which 
acts freely.  Thus, any naive attempt to wrap a p-brane on a torus would 
result in fixed-point singularities.  However, the antipodal map $J$ always
acts freely, and so one possible way to obtain a non-singular wrapped brane
is to compose the action of the lattice of spatial translations with the
involution $J$.

Of course, in situations involving several p-branes one often requires
the universal covering space of adS, denoted CadS, because several 
distinct adS patches may be required.  CadS is obtained by `unwrapping'
the time coordinate for adS; that is to say, on CadS the variable $t$ in
(3.4) is allowed to run over the whole real line.  One may then extend
the action of $J$ on $(adS)_{p+2}$ to a ${\Bbb Z}$-graded action on 
$(CadS)_{p+2}$ as shown \cite{gazza}:
\[
J^{n}: (t, {\chi}, {\Bbb n}) ~{\longrightarrow}~ (t + n{\pi}, {\chi}, (-1)^{n}{\Bbb n})
\]

With all of this in mind, let us now turn to the question of the existence
of Majorana pinors on the M-branes.  As we remarked above, it would appear
\cite{gazza} that whenever we wrap an M2 or M5 brane on a toroidal cycle,
we will have to simultaneously identify the adS factor of the throat
geometry of the brane if we want to avoid singularities.  Thus, before we
do anything we should check for the existence of Majorana fermions on
the orbifold M-branes.  Let's begin with the M5 brane.  As was pointed out
in \cite{gazza}, the action of $J$ on an odd-dimensional $(adS)_{p+2}$
is not only orientation preserving, but $J$ actually lies in the identity
component of the conformal group for $(p+1)$-dimensional Minkowski space.
It follows that ${w_{1}}^S(M;g_{L}) = 0$ and $w_{2}(M) = 0$ for the 
antipodally identified M5-brane, i.e., these branes in fact admit spin
structure so all of this concern about pin structure does not really apply
here.

The real issue is whether or not one can put Majorana structure on the 
antipodally identified M2-brane.  Indeed, if we antipodally identify the
$(adS)_4$ factor of the M2-brane, we obtain a non-space orientable manifold.
\footnote{Of course, if we require the existence of singletons then
we can only identify the covering space under the action of $J^4$, which
is orientation preserving \cite{gazza}.  Thus,  the existence of 
singletons will imply the existence of a spin structure.  Nevertheless,
the discussion presented here is relevant because it will apply in
any scenario where the M2-brane worldvolume is non-orientable, regardless
of whether or not there exist singletons or doubletons.}
In fact, the $S^2$-factor in the adS geometry is converted, under the
action of J, into a two-cycle with the topology of ${\Bbb R}{\Bbb P}^2$
(the `cross-cap').  $w_2 = 1$ on this two-cycle, and so this orbifold M2-brane
does {\it not} admit a spin structure.  On the other hand, one also calculates
that ${w_{1}}^S~{\smile}~{w_{1}}^S = 1$ on this two-cycle.  It follows
that the total obstruction ${\cal O}(M)$ actually vanishes, i.e., the
antipodally identified M2-brane does admit Majorana pinors.

This is reassuring, because it means we can wrap M-branes in space and time
without worrying about whether we might be selecting out the fermions which
are essential for the construction of the eleven-dimensional theory.
We can perform a similar analysis for membranes wrapped on a surface of
arbitrary genus \cite{rob}; there, one checks that cycles of arbitrary
genera admit the Majorana pin structure, i.e., ${\cal O}(M)$ vanishes
on any spacelike two-cycle, regardless of the genus and orientation.  

In general, the arguments presented here will go through for 
antipodally identified p-branes in any dimension, as long
as one is careful to choose a representation for which ${\cal P}^2 = -1$.
This sign ensures that the obstruction has the form (3.3), and hence
that the right sort of cancellation will occur for the non-orientable
branes.

This does not mean, however, that Majorana pinors are always allowed.
One can certainly imagine scenarios where one performs an exotic
orbifold projection on the transverse directions of a p-brane, or wraps
a p-brane around a cycle with an exotic topology, so that the resulting
spacetimes will not admit Majorana structure.  

Furthermore, it is worth pointing out that more stringent topological
conditions arise when one considers Type II string compactifications.
In particular, it has recently been shown \cite{witten} that in order
for a D-brane to consistently wrap a given cycle, the normal bundle
of the cycle must admit a ${\mbox{Spin}}^{c}$ structure.  The obstruction
theory for ${\mbox{Spin}}^{c}$ structures has been discussed recently in
\cite{bryant}, where it has been shown (among other things), that
the normal bundle for a SUSY cycle is generically ${\mbox{Spin}}^{c}$.

\section{Conclusions}

We have shown that it is possible to put Majorana pinors on M-branes
which are wrapped in space and in time.  If one chooses a natural,
Cliffordian choice for the representation of parity inversion then it
would seem that Majorana fermions select a unique pin structure.
In fact, given {\it any} choice of representation of ${\cal P}$,
the Majorana condition selects a unique pin group.  
This is because, once we have made a choice for the
representations of $P$ and $T$, we are not allowed to introduce any
complex numbers (this would violate the Majorana condition) and we are
not allowed to do any parity doubling (then the fermions would have the
wrong number of degrees of freedom for SUSY).  But these are the only 
two mechanisms which we can use to generate other representations for $P$
and $T$!  In other words, there is always only one choice of $P$ and $T$
consistent with the Majorana/SUSY conditions in eleven dimensions.
Does this uniqueness
perhaps imply that M-theory can solve the old problem of `the
classification of fermions'?

The problem of how one classifies fermions is simply this:
Does it make sense, or is it meaningful, to classify fermions 
according to their behaviour 
under the action of the full inhomogeneous Lorentz group? 
What would be the experimental consequences (if any) of such a classification
scheme?  Ever since Wigner \cite{wigner} introduced the different
corepresentations of $O(3,1)$ for the Dirac equation, people have
wondered about these things (see \cite{cec} for a modern viewpoint).

Suppose that we do classify fermions according to their behaviour
under the action of $P$ and $T$.  Then there are in principle eight
distinct particle `types', where the type is determined
by the pin group which acts on the fermion at a point of spacetime.
It is not hard to see that most of the observed elementary particles
can only come in one type.  For example, suppose that there existed two
types of electron, a `plus' type and a `minus' type.  The Pauli exclusion
principle would allow you to place a plus electron and a minus electron
in the same state.  Obviously, this would seriously mess up most of known
chemistry unless the electromagnetic interaction coupled only to
one type, and the other type was decoupled from known matter!
Thus, it would seem that nature has selected a particular pin structure
for the description of elementary particles.  From a four-dimensional
point of view, it is unclear why or how nature makes such a selection.

From the point of view of M-theory, however, the choice is obvious - 
Majorana selects a unique pin bundle.  Four-dimensional multiplets,
the descendants of the unique eleven-dimensional structure, then inherit
this choice.  This elegant solution of the classification problem is 
just another example of the power of M-theory.
\vspace*{0.3cm}

{\noindent \bf Acknowledgements}\\

It is a pleasure to thank G.W. Gibbons, N. Lambert 
and R. Mann for useful correspondence.
The author was supported by Pembroke College, University of Cambridge.

\bibliographystyle{board}
\bibliography{all}

\end{document}